\documentclass[reprint,twocolumn]{revtex4-2}
\usepackage{amsfonts}
\usepackage{amsmath}
\usepackage{amssymb}
\usepackage{charter}
\usepackage{graphicx}
\usepackage{float}
\usepackage{amsmath}
\usepackage{amssymb}
\usepackage{charter}
\usepackage{graphicx}
\usepackage{subfigure}
\usepackage{graphicx}
\usepackage{epstopdf}
\usepackage{color}
\usepackage{tocvsec2}
\usepackage{enumerate}
\usepackage{graphicx}
\usepackage{dcolumn}
\usepackage{bm}

\begin{document}

\title{ Improved phase sensitivity of an SU(1,1) interferometer based on the
internal single-path local squeezing operation }
\author{Qingqian Kang$^{1,2}$}
\author{Zekun Zhao$^{1}$}
\author{Teng Zhao$^{1}$}
\author{Cunjin Liu$^{1}$}
\author{Liyun Hu$^{1,3}$}
\thanks{hlyun@jxnu.edu.cn}
\affiliation{$^{{\small 1}}$\textit{Center for Quantum Science and Technology, Jiangxi
Normal University, Nanchang 330022, China}\\
$^{{\small 2}}$\textit{Department of Physics, College of Science and Technology,
Jiangxi Normal University, Nanchang 330022, China}\\
$^{{\small 3}}$\textit{Institute for Military-Civilian Integration of Jiangxi Province,
 Nanchang 330200, China}}

\begin{abstract}
Compared to passive interferometers, SU(1,1) interferometers exhibit
superior phase sensitivity due to the incorporation of nonlinear elements
that enhance their ability to detect phase shifts. However, the precision of
these interferometers is significantly affected by photon losses, especially
internal losses, which can limit the overall measurement accuracy.
Addressing these issues is essential to fully realize the advantages of
SU(1,1) interferometers in practical applications. Among the known resources
of quantum metrology, one of the most practical and efficient is squeezing.
We propose a theoretical scheme to improve the precision of phase
measurement using homodyne detection by implementing the single-path local
squeezing operation (LSO) inside the SU(1,1) interferometer, with the
coherent state and the vacuum state as the input states. We not only analyze
the effects of the single-path LSO scheme on the phase sensitivity and the
quantum Fisher information (QFI) under both ideal and photon-loss cases but
also compare the effects of different squeezing parameters $r$ on the system
performance. Our findings reveal that the internal single-path LSO scheme
can enhance the phase sensitivity and the QFI, effectively improving the
robustness of the SU(1,1) interferometer against internal and external
photon losses. Additionally, a larger squeezing parameter $r$ leads to a
better performance of the interferometer.

\end{abstract}

\maketitle

\section{Introduction}

Quantum precision measurement technology is a method that utilizes quantum
resources and effects to achieve measurement accuracy beyond traditional
methods. It integrates multidisciplinary knowledge such as atomic physics,
physical optics, electronic technology, and control technology, leveraging
principles of quantum mechanics, particularly the superposition and
entanglement properties of quantum states, to accomplish highly accurate
measurements. This technology has a wide range of applications \cite%
{2,3,4,5,6,7,8}, including high-precision optical frequency standards and
time-frequency transfer, quantum gyroscopes, atomic gravimeters, and other
quantum navigation technologies, as well as quantum radar, trace atom
tracking, and weak magnetic field detection in quantum sensitive detection
technologies. These technologies have significant application potential in
fields such as inertial navigation, next-generation time referencing,
stealth target identification, global terrain mapping, medical testing \cite%
{9}, and fundamental physics research \cite{10}.

Phase estimation is a crucial method for precision measurement, offering a
way to estimate many physical quantities that cannot be directly measured
through conventional methods. Consequently, extensive research and
significant advancements have been made in the field of optical interference
measurement. To satisfy the demand for high precision, various optical
interferometers have been proposed and developed. One of the most practical
interferometers is the Mach-Zehnder interferometer (MZI), whose phase
sensitivity is limited by the standard quantum-noise limit (SQL) $\Delta
\phi _{SQL}=1/\sqrt{N}$ ($N$ is the average number of photons within the
interferometer), together with solely classical resources as the input of
the MZI \cite{12}. In recent decades, various schemes have been proposed to
enhance the phase sensitivity of the traditional MZI \cite{b1,b2}. It has
been shown that using quantum states as input can enable the traditional MZI
to surpass the SQL. For instance, NOON states \cite{b3}, twin Fock states
\cite{b4}, squeezed states \cite{b5,b6}, and photon-catalyzed squeezed
vacuum states \cite{b7} can achieve or even exceed the Heisenberg limit (HL)
$\Delta \phi _{HL}=1/N$ \cite{b8}.

In 1986, Yurke \textit{et al.} \cite{a1} introduced the SU(1,1)
interferometer, which replaced traditional beam splitters (BSs) with optical
parametric amplifiers (OPAs). In the SU(1,1) interferometer consisting of
two OPAs, the first OPA serves the dual purpose of generating entangled
resources and suppressing amplified noise. The subsequent use of the second
OPA enhances the signal, providing a viable pathway to achieving higher
precision in phase estimation \cite{a2}. By utilizing entangled photon
states, the SU(1,1) interferometer can surpass the SQL, enabling greater
precision. This technique has revolutionized phase estimation, becoming a
vital tool in quantum precision measurements. In 2011, Jing \textit{et al.}
\cite{a3} successfully implemented this interferometer experimentally. For
instance, Hudelist \textit{et al.} demonstrated that the gain effect of OPAs
enables the SU(1,1) interferometer exhibiting higher sensitivity compared to
traditional linear interferometers \cite{a4}. This has led to a growing
interest in exploring the SU(1,1) interferometer \cite{a5,a6}. Apart from
the standard form, various configurations of SU(1,1) interferometer have
also been proposed \cite{a7,a8,a9,a10,a12,a13,a15,a16,a17}.

As previously mentioned, although SU(1,1) interferometer is highly valuable
for precision measurement \cite{c1,c1a}, the precision is still affected by
dissipation, particularly photon losses inside the interferometer \cite%
{c2,c3}. Consequently, to further enhance precision, non-Gaussian operations
can effectively mitigate internal dissipation. Most theoretical \cite%
{c5,c6,c61,c7} and experimental \cite{c8,c9,c10} studies have indicated that
non-Gaussian operations are effectively enhancing the nonclassicality and
entanglement degrees of quantum states, thereby enhancing their potential in
quantum information processing \cite{c11}.

It is evident that the use of non-Gaussian operations indeed improves the
estimation performance of optical interferometers, but this comes at a high
implementation cost. To address the aforementioned problem, Gaussian
operations such as the LSO \cite{m1,m2,m3} and the local displacement
operation (LDO) \cite{m4,m5} have emerged as promising strategies. In Ref.
\cite{m5}, Ye \textit{et al.} investigated phase sensitivity and QFI based
on homodyne detection in the presence and absence of photon losses through
displacement-assisted SU(1,1) [DSU(1,1)], involving two LDOs within the
SU(1,1) interferometer. In this DSU(1,1) interferometer, the introduced LDOs
improved phase sensitivity and QFI even under realistic conditions. It is
important to emphasize that the LSO is crucial not only in quantum metrology
\cite{m1} but also in quantum key distribution \cite{m2} and entanglement
distillation \cite{m3}. These applications demonstrate the versatility and
significance of the LSO in advancing various quantum technologies,
highlighting its role in enhancing precision measurements and ensuring
secure communication protocols. In Ref. \cite{m1}, J. Sahota and D. F. V.
James proposed an innovative quantum-enhanced phase estimation scheme that
applies the LSO to both paths of the MZI. Their work illustrates the
potential for the LSO scheme to enhance the performance of interferometric
devices, paving the way for advancements in quantum metrology and
measurement techniques.

Recent advances in the experimental realization of squeezing operations have
opened new avenues for enhancing the precision of optical phase measurements
\cite{n1,n2,n3,n4}. Thus, can the single-path LSO scheme implemented inside
the SU(1,1) interferometer similarly enhance the phase sensitivity and the
QFI? Can this approach enhance the robustness of the interferometer against
internal and external photon losses? If the single-path LSO scheme brings
improvements, it would be a better choice from the perspective of quantum
resources.

In this paper, we propose a theoretical scheme to improve the precision of
phase measurement using homodyne detection by implementing the single-path
LSO scheme inside the SU(1,1) interferometer, with the coherent state and
the vacuum state as the input states. We also perform a comparative analysis
about the effects of internal and external photon losses on interferometer
performance. This paper is arranged as follows. Sec. II outlines the
theoretical model of the SU(1,1) interferometer based on the internal
single-path LSO scheme. Sec. III delves into phase sensitivity, encompassing
both ideal and photon-loss cases. Sec. IV centers on the QFI. Finally, Sec.
V provides a summary.

\section{Model}

This section first describes the standard SU(1,1) interferometer, as shown
in Fig. 1(a). The SU(1,1) interferometer, which usually consists of two OPAs
and a linear phase shifter, is one of the most commonly used interferometers
in quantum metrology studies. The first OPA is characterized by a two-mode
squeezing operator $U_{S_{1}}(\xi _{1})=\exp (\xi _{1}^{\ast }ab-\xi
_{1}a^{\dagger }b^{\dagger })$, where $a$ ($b$) and $a^{\dagger }$ ($%
b^{\dagger }$) represent the photon annihilation and creation operators,
respectively. The squeezing parameter $\xi _{1}$ can be expressed as $\xi
_{1}=g_{1}e^{i\theta _{1}}$, where $g_{1}$ represents the gain factor and $%
\theta _{1}$ represents the phase shift. This parameter plays a critical
role in shaping the interference pattern and determining the system's phase
sensitivity. After the first OPA, mode $a$ undergoes a phase shift process $%
U_{\phi }=\exp [-i\phi (a^{\dagger }a)]$, while mode $b$ remains unchanged.
Subsequently, the two beams are coupled in the second OPA with the operator $%
U_{S_{2}}(\xi _{2})=\exp (\xi _{2}^{\ast }ab-\xi _{2}a^{\dagger }b^{\dagger
})$, where $\xi _{2}=g_{2}e^{i\theta _{2}}$ with $\theta _{2}-\theta
_{1}=\pi $. In this paper, we set the parameters $g_{1}=g_{2}=g$, $\theta
_{1}=0$, and $\theta _{2}=\pi $. We use the coherent state $\left \vert
\alpha \right
\rangle _{a}$ and the vacuum state $\left \vert
0\right
\rangle _{b}$ as input states, and homodyne detection is applied to
the mode $a$ of the output.

\begin{figure}[tph]
\label{Figure1} \centering \includegraphics[width=0.85\columnwidth]{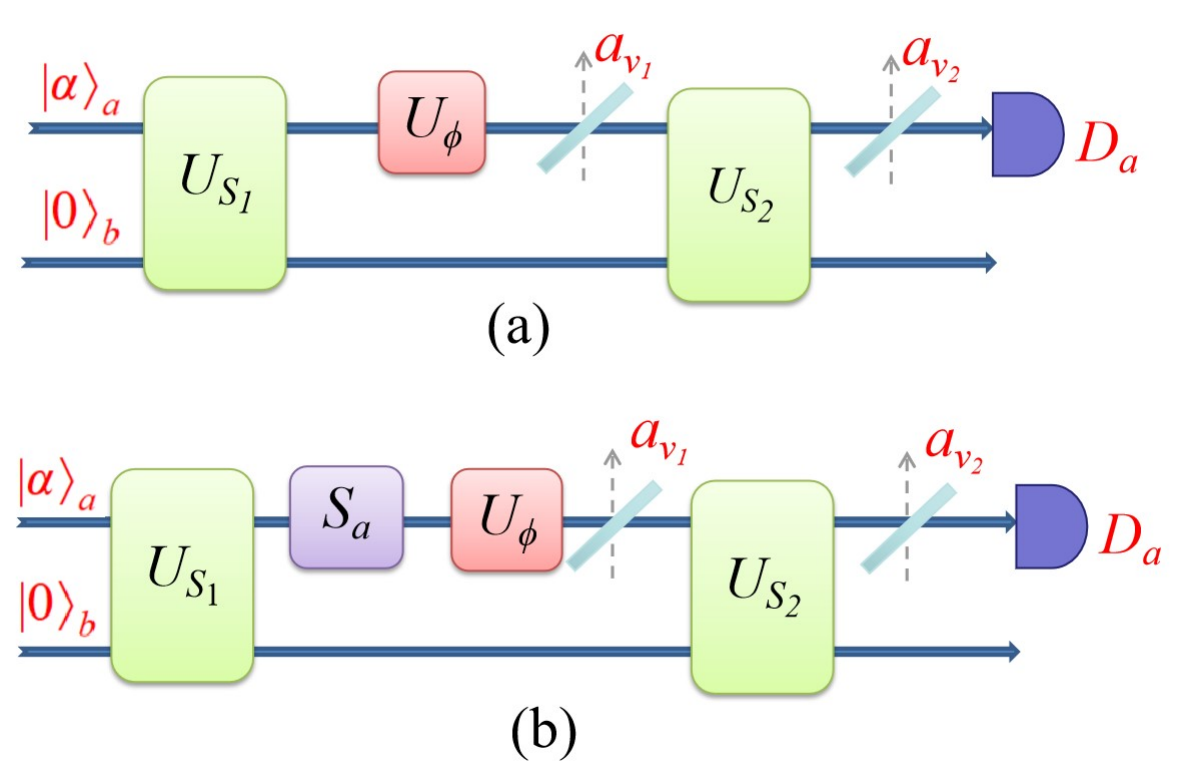}
\caption{Schematic diagram of the SU(1,1) interferometer. (a) The standard
SU(1,1) interferometer and (b) the SU(1,1) interferometer with the
single-path LSO scheme. The two input ports are a coherent state $%
\left
\vert \protect \alpha \right
\rangle _{a}$ and a vacuum state $%
\left
\vert 0\right \rangle _{b}$. $a_{v_{1}}$ and $a_{v_{2}}$ are vacuum
modes. $U_{S_{1}}$ and $U_{S_{2}}$\ are the OPAs, and $U_{\protect \phi }$ is
the phase shifter. $S_{a}$ is the single-path LSO operator and $D_{a}$ is
the homodyne detector.}
\end{figure}

The SU(1,1) interferometer is generally susceptible to photon losses,
particularly in the case of internal losses. To simulate internal and
external photon losses, the use of fictitious BSs is proposed, as depicted
in Fig. 1(a). The operators of these fictitious BSs can be represented as $%
U_{B_{1}}\ $and $U_{B_{2}}$, with $U_{B_{1}}=\exp \left[ \theta
_{T_{1}}\left( a^{\dagger }a_{v_{1}}-aa_{v_{1}}^{\dagger }\right) \right] $
and $U_{B_{2}}=\exp \left[ \theta _{T_{2}}\left( a^{\dagger
}a_{v_{2}}-aa_{v_{2}}^{\dagger }\right) \right] $, where $a_{v_{1}}$ and $%
a_{v_{2}}$ represent vacuum modes. Here, $T_{k}$ ($k=1,2$) denotes the
transmissivity of the fictitious BSs, associated with $\theta _{T_{k}}$
through $T_{k}=\cos ^{2}\theta _{T_{k}}\in \left[ 0,1\right] $. The value of
transmittance equal to $1$ ($T_{k}=1$) corresponds to the ideal case without
photon losses \cite{c13}. In an expanded space, the expression for the
output state of the standard SU(1,1) interferometer can be represented as
the following pure state:%
\begin{equation}
\left \vert \Psi _{out}^{0}\right \rangle
=U_{B_{2}}U_{S_{2}}U_{B_{1}}U_{\phi }U_{S_{1}}|\psi _{in}\rangle ,
\label{eq1}
\end{equation}%
where $|\psi _{in}\rangle =$ $\left \vert \alpha \right \rangle
_{a}\left
\vert 0\right \rangle _{b}\left \vert 0\right \rangle
_{a_{v_{1}}}\left
\vert 0\right \rangle _{a_{v_{2}}}$.

To mitigate the impact of photon losses, we introduce a Gaussian operation
inside the SU(1,1) interferometer, called the single-path LSO scheme, as
illustrated in Fig. 1(b). Recent advancements in the experimental
realization of squeezing operations has significantly contributed to the
field of quantum metrology, particularly in enhancing the sensitivity of
interferometers beyond the SQL. For instance, Purdy \textit{et al.} \cite{n1}
experimentally demonstrated strong and continuous optomechanical squeezing
of $1.7\pm 0.2$\ dB below the shot-noise level by exploiting the quantum
interaction between laser light and a membrane mechanical resonator in an
optical cavity. Meanwhile, Ono \textit{et al.} \cite{n2} explored optical
spin-squeezing, revealing a quantum advantage of 1.58 over the shot-noise
limit for five-photon events, suggesting enhanced performance for quantum
metrology applications, despite not achieving sub-shot-noise precision due
to experimental imperfections. Furthermore, Zuo \textit{et al.} \cite{n3}
proposed and experimentally demonstrated a compact quantum interferometer
that achieves a sensitivity improvement of $4.86\pm 0.24$\ dB beyond the SQL
by using squeezed states generated within the MZI for phase sensing,
resulting in a minimum detectable phase smaller than that of all present
interferometers under the same phase-sensing intensity. In 2023, Kalinin
\textit{et al.} \cite{n4} reported the first experimental demonstration of
phase sensitivity enhancement in an interferometer using Kerr squeezing,
addressing the challenge of the cumbersome tilting of the squeezed ellipse
in phase space. The experimental realization of squeezing operations has
advanced significantly, enabling the development of highly sensitive and
robust interferometric schemes that surpass classical limits. These
advancements hold promise for a wide range of applications, including
gravitational wave detection, quantum communication, and precision sensing.

In our scheme, we utilize simple and easy-to-prepare input states ($\left
\vert \alpha \right \rangle _{a}\otimes \left \vert 0\right \rangle _{b}$),
and an experimentally feasible homodyne detection. The operator of the
single-path LSO inside the SU(1,1) interferometer on mode $a$ can be written
as:
\begin{equation}
S_{a}=\exp [\frac{1}{2}\left( re^{-i\pi }a^{2}-re^{i\pi }a^{\dagger
2}\right) ].  \label{eq3}
\end{equation}%
In this case of the single-path LSO scheme applied inside the SU(1,1)
interferometer, the output state can be written as the following pure state:
\begin{equation}
\left \vert \Psi _{out}^{1}\right \rangle
=U_{B_{2}}U_{S_{2}}U_{B_{1}}U_{\phi }S_{a}U_{S_{1}}|\psi _{in}\rangle .
\label{eq4}
\end{equation}

\section{Phase sensitivity}

Quantum metrology utilizes quantum resources to achieve precise phase
measurements \cite{d1,d2}. The primary objective is to attain highly
sensitive measurements of unknown phases. Phase sensitivity is a crucial
parameter because improving it can reduce the uncertainty in phase
measurements, thereby enhancing the accuracy of the measurements \cite{d3}.
Selecting an appropriate detection method to extract the phase information
is the crucial final step in the phase estimation process. This step is
essential for ensuring the accuracy and reliability of measurement results.
Common detection methods include homodyne detection \cite{d4,d5}, parity
detection \cite{b7,d7}, and intensity detection \cite{d8}, each offering
unique trade-offs among sensitivity, complexity, and practical application.
It is important to note that the phase sensitivity of different detection
schemes can vary depending on the input state and the design of the
interferometerd \cite{d9}. Homodyne detection is feasible with current
experimental techniques \cite{c15a}, and its theoretical calculations are
relatively straightforward. Therefore, we opt to use homodyne detection at
output port $a$ to estimate phase sensitivity.

In homodyne detection, the measured variable is one of the two orthogonal
components of mode $a$, given by $X=(a+a^{\dagger })/\sqrt{2}$. Based on the
error-propagation equation \cite{a1}, the phase sensitivity can be expressed
as
\begin{equation}
\Delta \phi =\frac{\sqrt{\left \langle \Delta ^{2}X\right \rangle }}{%
|\partial \left \langle X\right \rangle /\partial \phi |}=\frac{\sqrt{\left
\langle X^{2}\right \rangle -\left \langle X\right \rangle ^{2}}}{|\partial
\left \langle X\right \rangle /\partial \phi |}.  \label{eq12}
\end{equation}

Based on Eqs. (\ref{eq4}) and (\ref{eq12}), the phase sensitivity for the
LSO scheme can be theoretically determined. Detailed calculation steps for
the phase sensitivity $\Delta \phi $ of the single-path LSO scheme are
provided in Appendix A.

\subsection{Ideal case}

Initially, we consider the ideal case, $T_{k}=1$ (where $k=1,2$),
representing the case without photon losses. The phase sensitivity $\Delta
\phi $ is plotted as a function of $\phi $ for various squeezing parameters $%
r$ in Fig. 2. It is shown that: (i) The phase sensitivity improves initially
and then decreases as the phase $\phi $ increases, with the optimal
sensitivity deviating from $\phi =0$. (ii) The implementation of the
single-path LSO inside the SU(1,1) interferometer significantly enhances the
phase sensitivity $\Delta \phi $. This enhancement becomes especially
pronounced with higher squeezing parameters. (iii) As the value of $r$
increases, the optimal phase sensitivity moves away from $\phi =0$.

\begin{figure}[tph]
\label{Figure2} \centering \includegraphics[width=0.85\columnwidth]{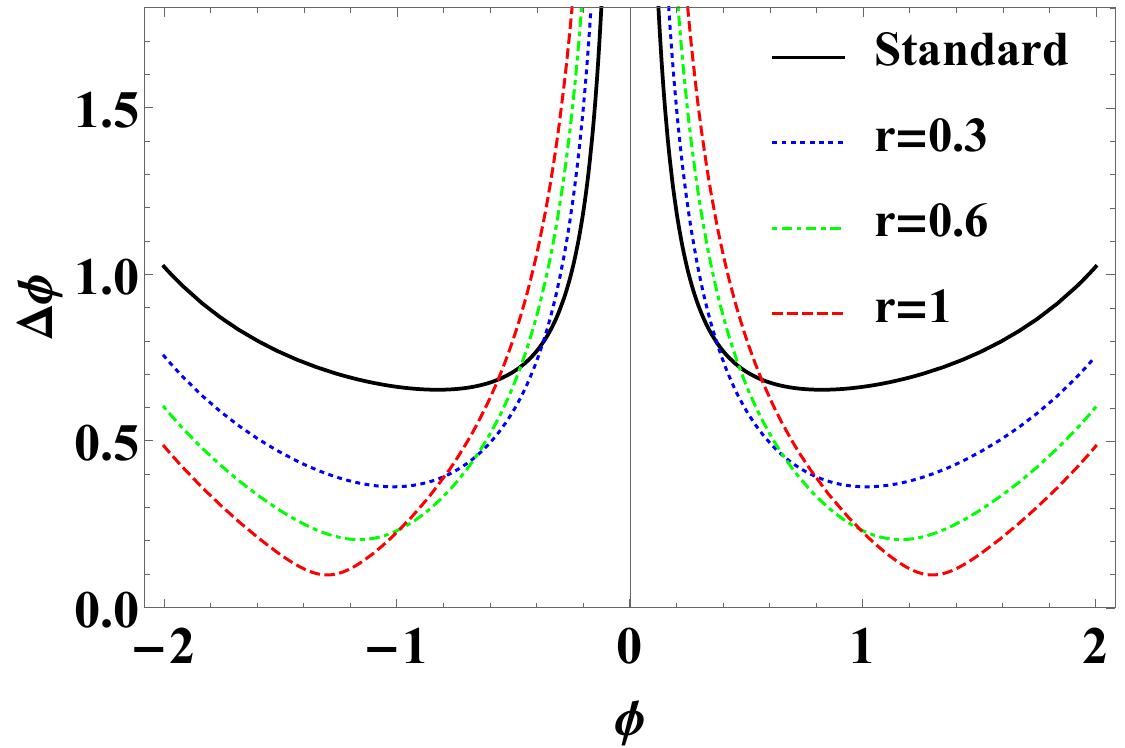}
\caption{The phase sensitivity of single-path LSO based on the homodyne
detection as a function of $\protect \phi $ with $\protect \alpha =1$ and $g=1$%
. The black solid line represents the standard SU(1,1) interferometer, while
the blue dotted line, green dot-dash line, and red dashed line correspond to
squeezing parameters of $r=0.3$, $r=0.6$ and $r=1$, respectively.}
\end{figure}

Fig. 3 illustrates that the optimal phase sensitivity (minimizing $\Delta
\phi $) is plotted against the gain factor $g$ for different squeezing
parameters. The plot confirms that the single-path LSO scheme enhances the
phase sensitivity $\Delta \phi $. The improvement effect changes little with
increasing $g$. And, the larger the squeezing parameter $r$, the better the
phase sensitivity $\Delta \phi $.

\begin{figure}[tph]
\label{Figure3} \centering \includegraphics[width=0.85\columnwidth]{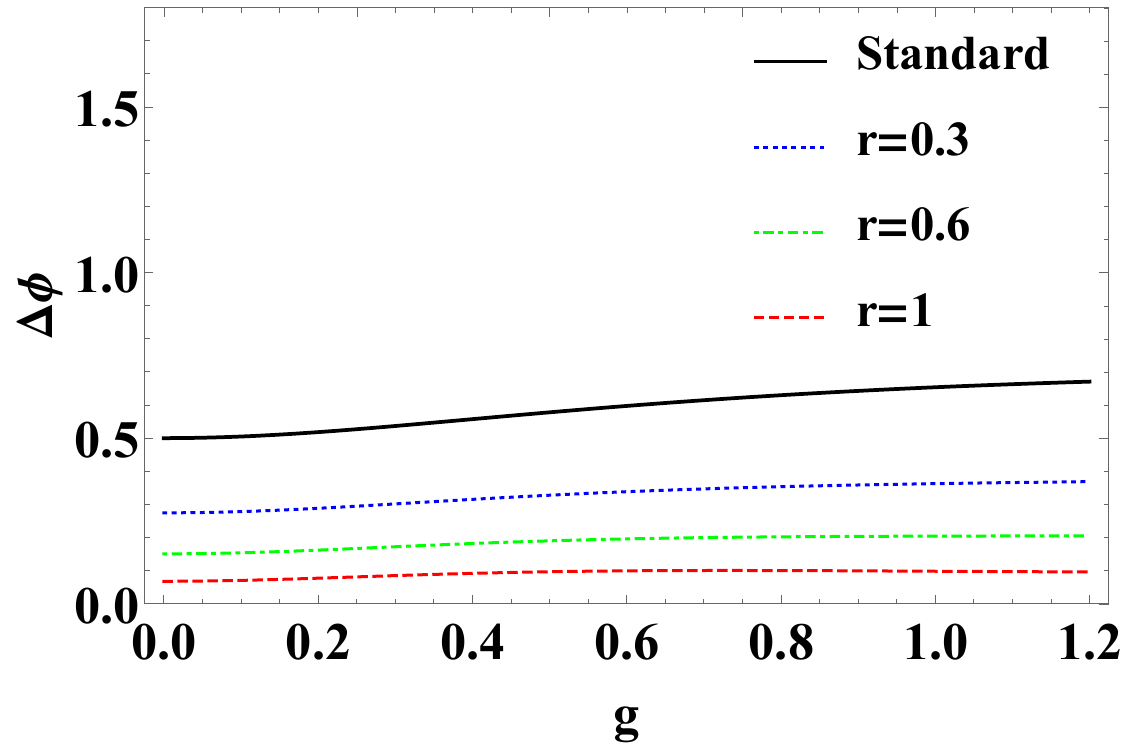}
\caption{The phase sensitivity as a function of $g$, with $\protect \alpha %
=1\ $. }
\end{figure}

Similarly, we analyze the optimal phase sensitivity (minimizing $\Delta \phi
$) as a function of the coherent amplitude $\alpha $, as depicted in Fig. 4.
The phase sensitivity improves with the coherent amplitude $\alpha $,
attributed to the increase in the average photon number with $\alpha $, then
enhancing intramode correlations \cite{c19} and quantum entanglement between
the two modes. Furthermore, the enhancement effect diminishes as the
coherent amplitude $\alpha $ increases. Again, the larger the squeezing
parameter $r$, the better the phase sensitivity $\Delta \phi $.

\begin{figure}[tph]
\label{Figure4} \centering \includegraphics[width=0.85\columnwidth]{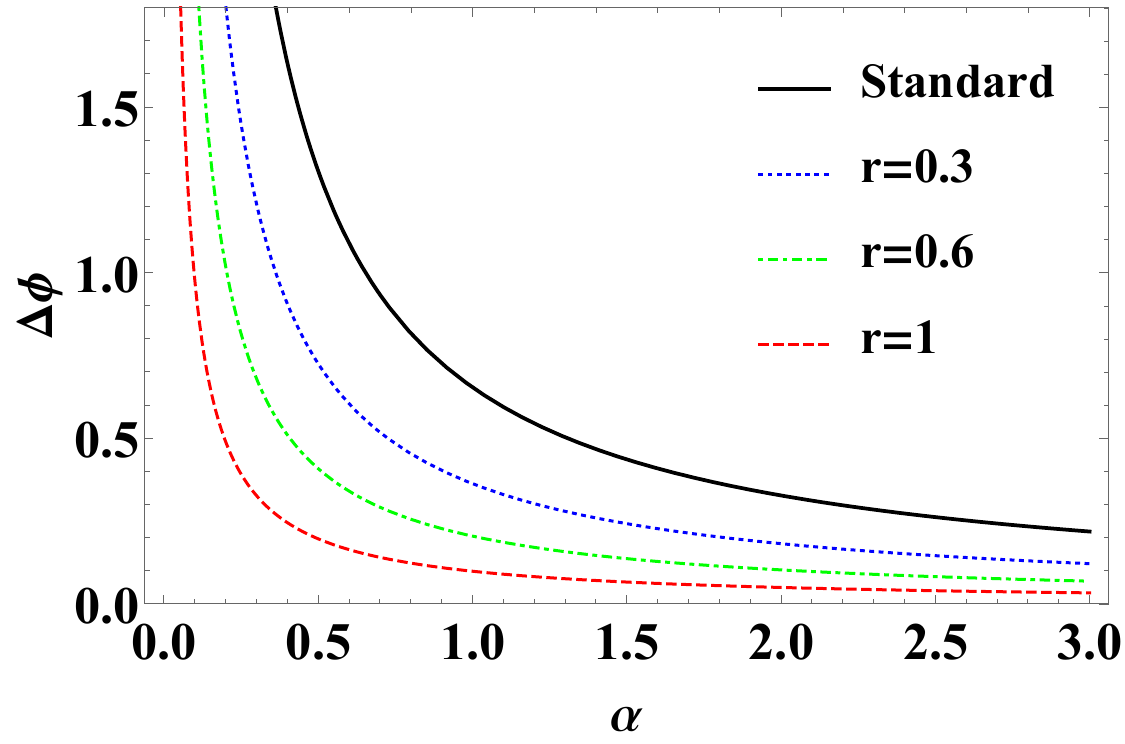}
\caption{The phase sensitivity as a function of $\protect \alpha $, with $%
g=1\ $. }
\end{figure}

\subsection{Photon-loss case}

The SU(1,1) interferometer plays a critical role in achieving high-precision
measurements, but the measurement accuracy is highly sensitive to photon
losses. Here, we focus on internal and external photon losses, corresponding
to $T_{k}\in (0,1)$. The optimal phase sensitivity (minimizing $\Delta \phi $%
), depicted as a function of transmittance $T_{k}$ in Fig. 5 for fixed $g$, $%
\alpha $, and $\phi $, improves as anticipated with higher transmittance $%
T_{k}$. Lower transmittance corresponds to increased losses, weakening the
performance of phase estimation. Both internal and external photon losses
degrade phase sensitivity. Notably, throughout the entire range of $T_{k}$,
internal losses have a more significant impact on the system's phase
sensitivity. This is primarily because the second OPA amplifies both the
signal and the internal noise. The improved effects of phase sensitivity are
also clearly observed with an increase in the squeezing parameter $r$. In
contrast, the phase sensitivity of the single-path LSO scheme is less
affected, indicating that the Gaussian operations can mitigate the impact of
internal or external photon losses and enhance the interferometer's
robustness against losses.

\begin{figure}[tph]
\label{Figure5} \centering \includegraphics[width=0.85\columnwidth]{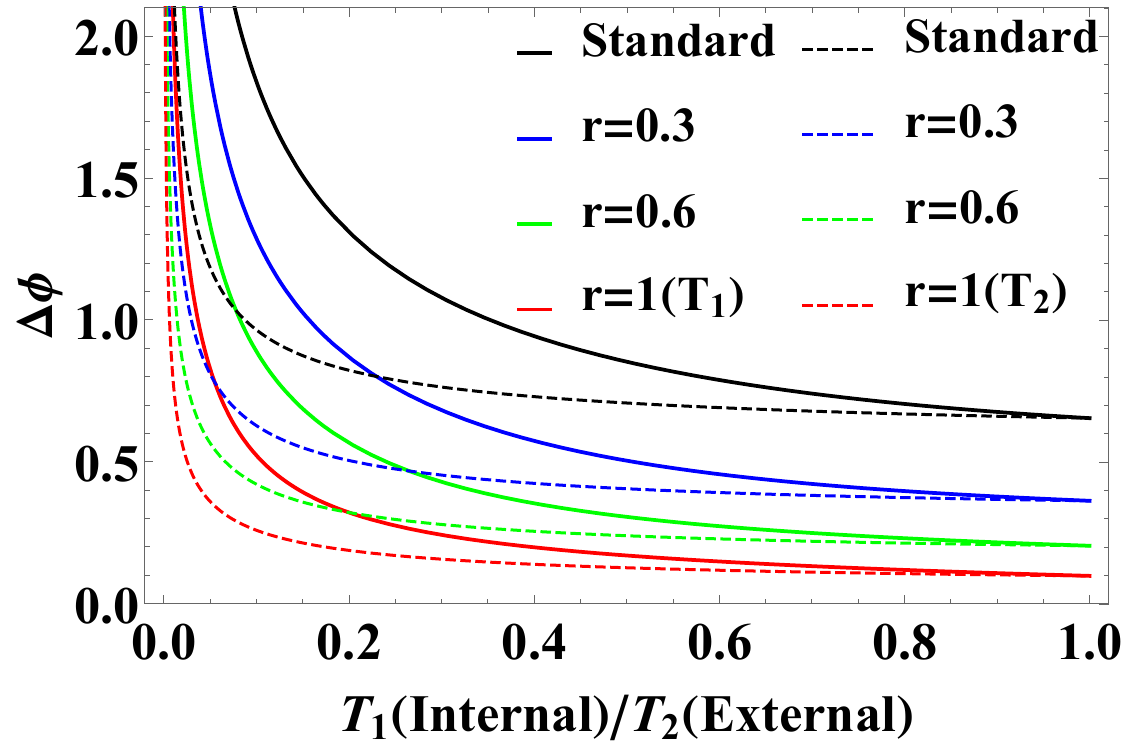}
\caption{The phase sensitivity as a function of transmittance $T_{k}$, with $%
g=1$ and $\protect \alpha =1.$ The solid lines correspond to internal losses,
while the dashed lines correspond to external losses.}
\end{figure}

\subsection{Comparison with SQL and HL}

Additionally, we compare the phase sensitivity with the SQL and HL in this
section. The SQL and HL are defined as $\Delta \phi _{SQL}=1/\sqrt{N}$ and $%
\Delta \phi _{HL}=1/N$, respectively. Here $N$ represents the total average
photon number inside the interferometer before the second OPA for the ideal
case \cite{12,e3}. $N$ can be calculated as:
\begin{eqnarray}
N &=&\langle \psi _{in}|U_{S_{1}}^{\dagger }S_{a}^{\dagger }\left(
a^{\dagger }a+b^{\dagger }b\right) S_{a}U_{S_{1}}|\psi _{in}\rangle  \notag
\\
&=&Q_{1,1,0,0}+Q_{0,0,1,1},  \label{eq13}
\end{eqnarray}%
where the expression of $Q_{x_{1},y_{1},x_{2},y_{2}}$ is given in Appendix A.

\begin{figure}[tph]
\label{Figure6} \centering%
\subfigure{
\begin{minipage}[b]{0.5\textwidth}
\includegraphics[width=0.85\textwidth]{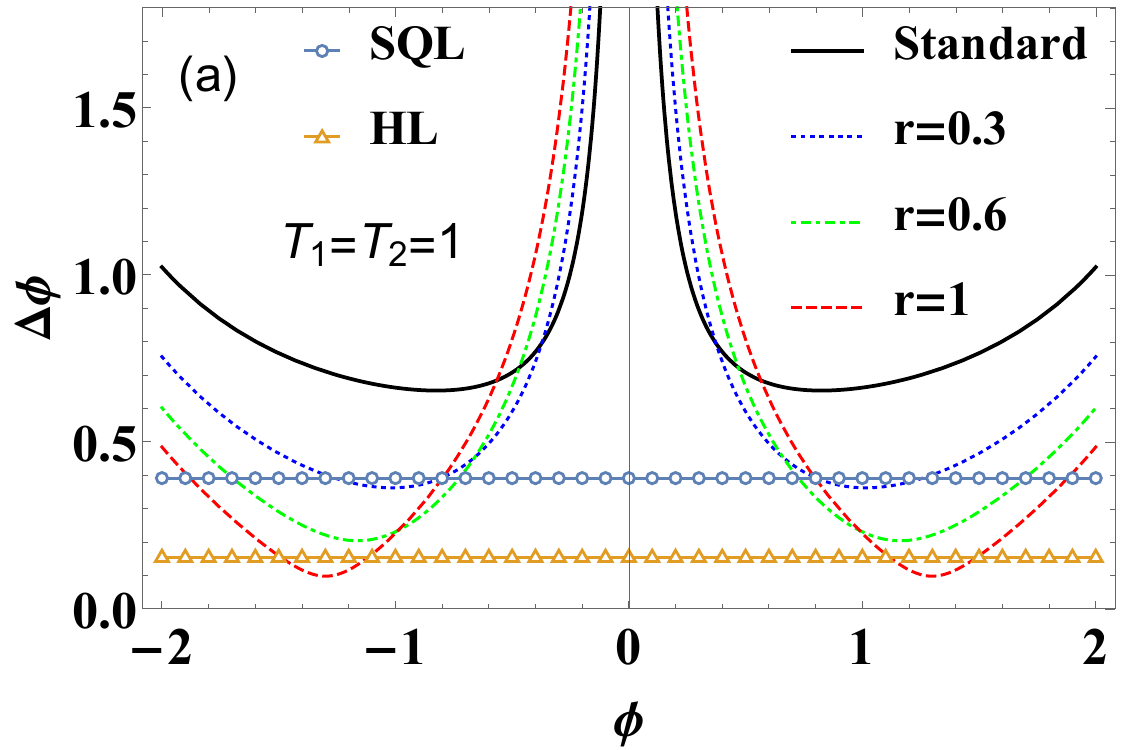}\\
\includegraphics[width=0.85\textwidth]{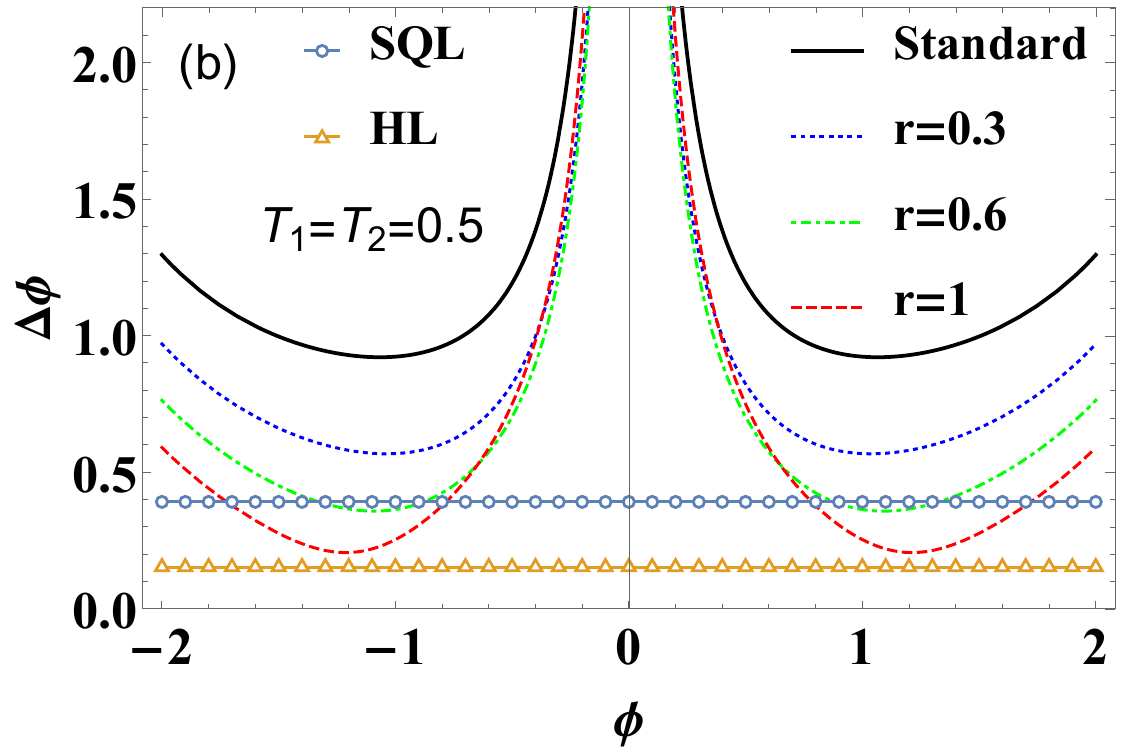}
\end{minipage}}
\caption{Comparison of phase sensitivity with the SQL and the HL for fixed $%
g=1$ and $\protect \alpha =1$. The blue circle is the SQL and the yellow
triangle is the HL. (a) $T_{1}=T_{2}=1$ and (b) $T_{1}=T_{2}=0.5$.}
\end{figure}

For fixed $\alpha $ and $g$, we plot the phase sensitivity $\Delta \phi $ as
a function of $\phi $ for a comparison with the SQL and the HL of the
standard SU(1,1) interferometer in Fig. 6. Our findings demonstrate that:
(i) In the ideal case, the phase sensitivity of the standard interferometer
(without LSO) cannot surpass the SQL and the HL. However, with the
single-path LSO scheme, it can surpass the SQL even when the squeezing
parameter $r$ is relatively small ( $r=0.3$). Moreover, as $r$ increases,
the sensitivity of the LSO scheme can surpass the HL. (ii) In the presence
of significant photon losses, the single-path LSO scheme still demonstrates
a capability to surpass the SQL when the squeezing parameter $r$ is large.
As $r$ increases, the single-path LSO scheme approaches the HL. This
indicates that the single-path LSO scheme exhibits strong robustness against
photon losses, effectively addressing the challenges posed by photon losses
in practical applications.

\section{The quantum Fisher information}

In our previous discussions, we investigated how the single-path LSO scheme
affects phase sensitivity and the relationship between various relevant
parameters and phase sensitivity through homodyne detection. It is important
to note that the phase sensitivity is influenced by the specific measurement
method employed. This leads us to the critical question: how can we attain
the highest phase sensitivity in an interferometer that is unaffected by the
choice of measurement method? In this section, we will turn our attention to
the QFI, which indicates the maximum amount of information that can be
extracted from the interferometer system, irrespective of the measurement
technique used. We will analyze the QFI under both ideal and realistic
conditions.

\subsection{Ideal case}

For a pure state system, the QFI can be derived by \cite{e4}%
\begin{equation}
F=4\left[ \left \langle \Psi ^{\prime }|\Psi ^{\prime }\right \rangle -\left
\vert \left \langle \Psi ^{\prime }|\Psi \right \rangle \right \vert ^{2}%
\right] ,  \label{eq15}
\end{equation}%
where\ $\left \vert \Psi \right \rangle $ is the quantum state after phase
shift and before the second OPA, and $\left \vert \Psi ^{\prime
}\right
\rangle =\partial \left \vert \Psi \right \rangle /\partial \phi .$
Then the QFI can be reformed as \cite{e4}:
\begin{equation}
F=4\left \langle \Delta ^{2}n_{a}\right \rangle ,  \label{eq16}
\end{equation}%
where $\left \langle \Delta ^{2}n_{a}\right \rangle =\left \langle \Psi
\right \vert (a^{\dagger }a)^{2}|\Psi \rangle -\left( \left \langle \Psi
\right \vert a^{\dagger }a|\Psi \rangle \right) ^{2}$.

In the ideal case, the quantum state is given by $\left \vert \Psi
\right
\rangle =U_{\phi }S_{a}U_{S_{1}}\left \vert \alpha \right \rangle
_{a}\left
\vert 0\right \rangle _{b}$. Thus, the QFI is derived as:
\begin{equation}
F=4\left( Q_{2,2,0,0}+Q_{1,1,0,0}\right) -4\left( Q_{1,1,0,0}\right) ^{2}%
\text{,}  \label{eq17}
\end{equation}%
where the expression for $Q_{x_{1},y_{1},x_{2},y_{2}}$ is given in Appendix
A. It is possible to explore the connection between the QFI and the related
parameters using Eq. (\ref{eq17}).

Fig. 7 illustrates the QFI as a function of $g$ ($\alpha $) for a specific $%
\alpha $ ($g$). It is evident that a higher value of $g$ ($\alpha $)
corresponds to a greater QFI. The larger the squeezing parameter $r$, the
greater the QFI. Moreover, we observe that the improvement in QFI due to
Gaussian operation increases with the increase of the value $g$ ($\alpha $).

\begin{figure}[tph]
\label{Figure7} \centering%
\subfigure{
\begin{minipage}[b]{0.5\textwidth}
\includegraphics[width=0.85\textwidth]{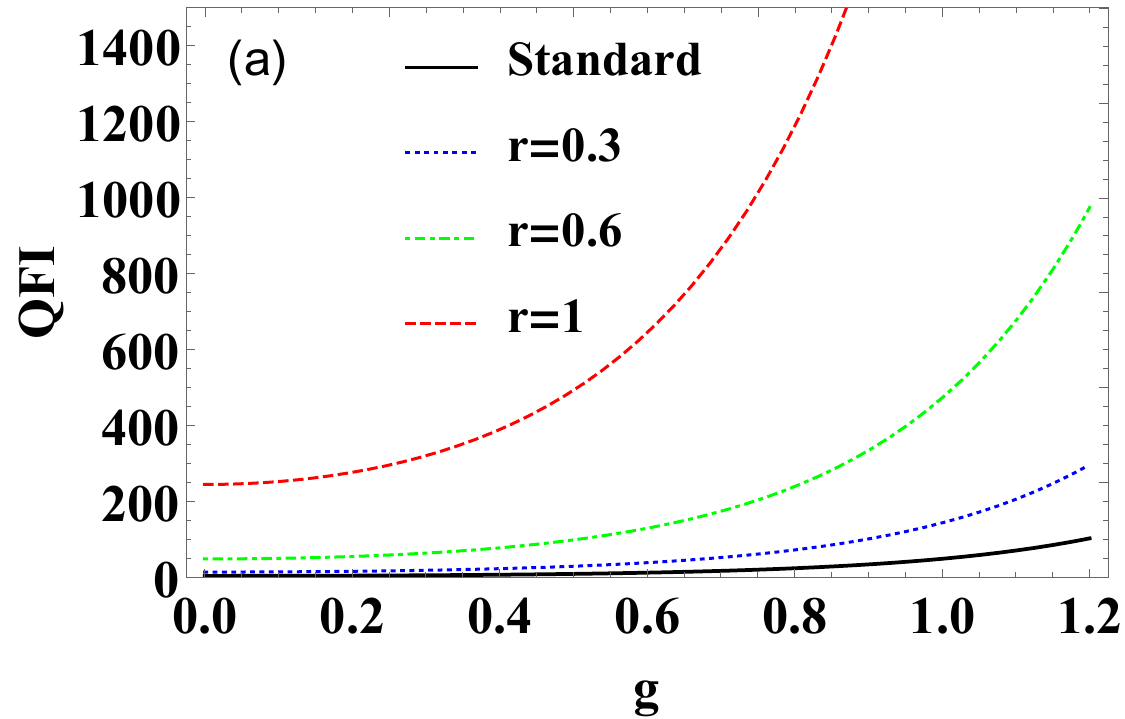}\\
\includegraphics[width=0.85\textwidth]{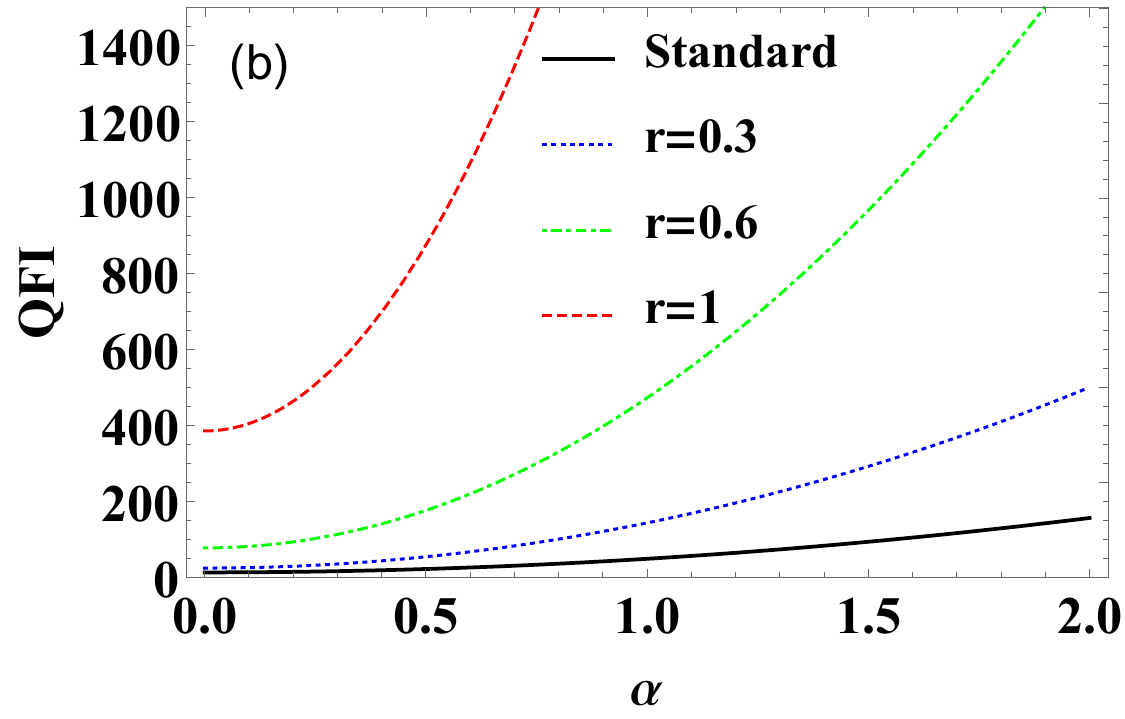}
\end{minipage}}
\caption{(a) The QFI as a function of $g$, with $\protect \alpha =1$. (b) The
QFI as a function of $\protect \alpha $, with $g=1$. }
\end{figure}

The minimum value of phase sensitivity achievable for all measurement
schemes is known as quantum Cram\'{e}r-Rao bound (QCRB) as defined by \cite%
{c14}:%
\begin{equation}
\Delta \phi _{QCRB}=\frac{1}{\sqrt{vF}},  \label{eq19}
\end{equation}%
where $v$ represents the number of measurements. For simplicity, we set $v=1$%
. The QCRB \cite{c14,c15}, denoted as $\Delta \phi _{QCRB}$, establishes the
ultimate limit for a set of probabilities derived from measurements on a
quantum system. It serves as an estimator implemented asymptotically via
maximum likelihood estimation and provides a measurement-independent phase
sensitivity. To evaluate the optimality of the phase sensitivity achieved by
the SU(1,1) interferometer with the single-path LSO scheme, we analyze the
sensitivity $\Delta \phi _{QCRB}$ derived from the QFI in Fig. 8. It
illustrates the variation of $\Delta \phi _{QCRB}$ as a function of $g$ ($%
\alpha $) for a specific $\alpha $ ($g$). It is shown that $\Delta \phi
_{QCRB}$ improves with increasing $g$ and $\alpha $. Similarly, the larger
the squeezing parameter $r$, the better the $\Delta \phi _{QCRB}$.
Furthermore, the improvement in $\Delta \phi _{QCRB}$ is more obvious for
small gain factor $g$ [refer to Fig. 8(a)] and small coherent amplitude $%
\alpha $ [refer to Fig. 8(b)].

\begin{figure}[tph]
\label{Figure8} \centering%
\subfigure{
\begin{minipage}[b]{0.5\textwidth}
\includegraphics[width=0.85\textwidth]{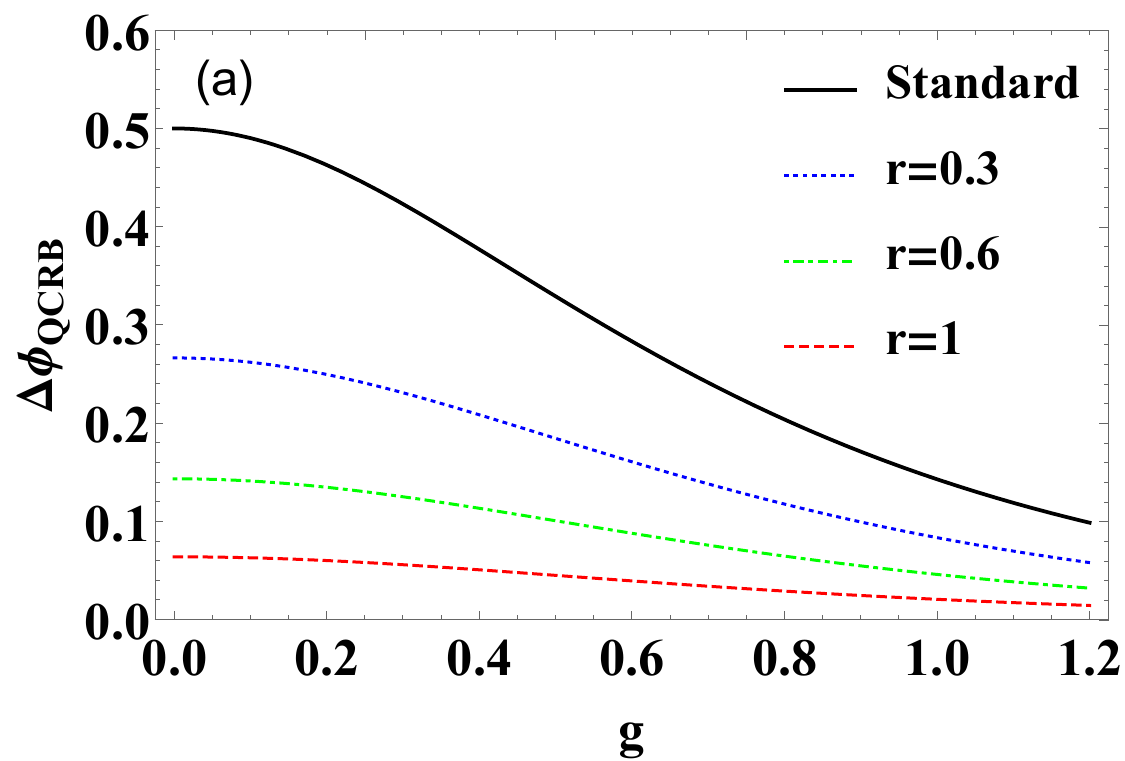}\\
\includegraphics[width=0.85\textwidth]{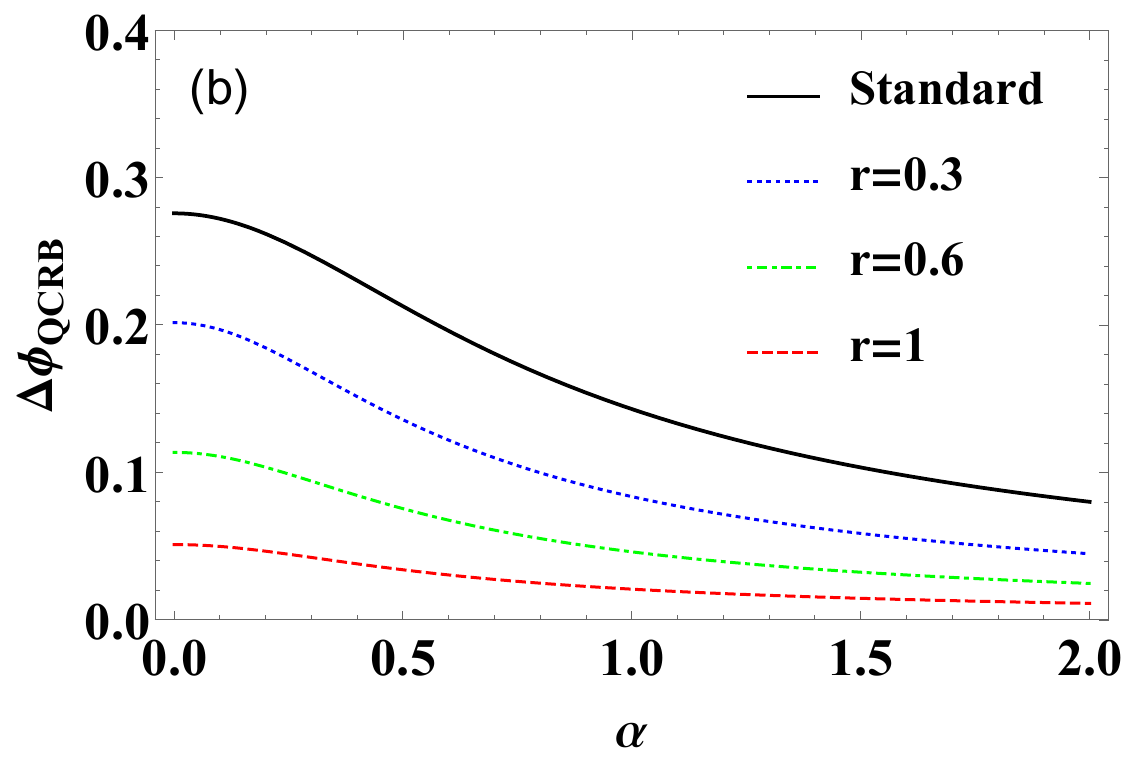}
\end{minipage}}
\caption{(a) The $\Delta \protect \phi _{QCRB}$ as a function of $g$, with $%
\protect \alpha =1$. (b) The $\Delta \protect \phi _{QCRB}$ as a function of $%
\protect \alpha $, with $g=1$. }
\end{figure}

\subsection{Photon losses case}

In this subsection, we extend our analysis to cover the QFI in the presence
of photon losses. Specifically, we examine homodyne detection on mode $a$,
which is susceptible to photon losses. Consequently, our attention is
directed toward the QFI of the system with photon losses on mode $a$, as
depicted in Fig. 9. Here, we should emphasize that the Fisher information is
obtained using the state preceding the second OPA, i.e., the second OPA is
not essential. For realistic quantum systems, Escher \textit{et al}. \cite%
{e4} propose a method for calculating the QFI. This method can be briefly
summarized as follows.

\begin{figure}[tph]
\label{Figure9} \centering \includegraphics[width=0.85\columnwidth]{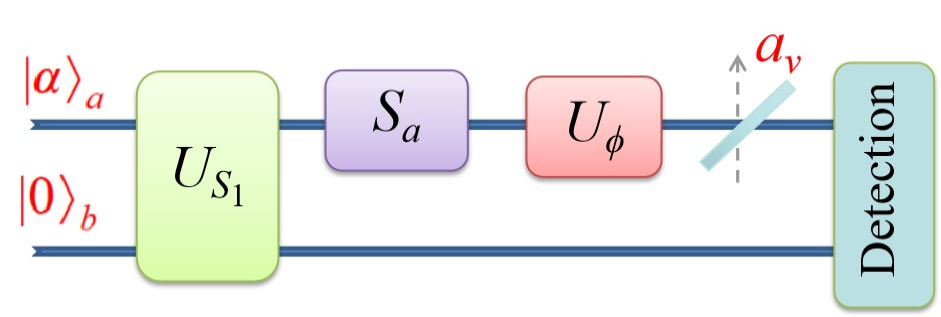}
\
\caption{Schematic diagram of the photon losses on mode $a$. The losses
occurs after the $U_{\protect \phi }$.}
\end{figure}

The QFI with photon losses is calculated as detailed in Ref. \cite{e4}.
After the first OPA $U_{S_{1}}$, the Gaussian operation $S_{a}$, the phase
shift $U_{\phi }$, the photon losses $U_{B}$, and before the detection, the
output state in an expanded space can be given by
\begin{equation}
\left \vert \Psi _{S}\right \rangle =U_{B}U_{\phi }S_{a}\left \vert 0\right
\rangle _{a_{v}}\left \vert \psi \right \rangle ,  \label{eq20}
\end{equation}%
a form of pure state, where $\left \vert \psi \right \rangle
=U_{S_{1}}\left
\vert \alpha \right \rangle _{a}\left \vert 0\right \rangle
_{b}$.

For the case of photon losses, we can treat the system as a pure state in an
extended space, similar to Eq. (\ref{eq4}). Then following Eq. (\ref{eq15}),
we can obtain the QFI under the pure state, denoted as $C_{Q}$, which is
larger or equal to the QFI for mixed state, denoted as $F_{L}$, i.e., $%
F_{L}\leq C_{Q}$. $C_{Q}$ is the QFI before optimizing over all possible
measurements, i.e.,%
\begin{equation}
C_{Q}=4\left[ \left \langle \psi \right \vert H_{1}\left \vert \psi \right
\rangle -\left \vert \left \langle \psi \right \vert H_{2}\left \vert \psi
\right \rangle \right \vert ^{2}\right] ,  \label{eq21}
\end{equation}%
where $H_{1}$ and $H_{2}$ are defined as%
\begin{eqnarray}
H_{1} &=&\overset{\infty }{\underset{l=0}{\sum }}\frac{d}{d\phi }\Pi
_{l}^{\dagger }\left( \eta ,\phi ,\lambda \right) \frac{d}{d\phi }\Pi
_{l}\left( \eta ,\phi ,\lambda \right) ,  \label{eq22} \\
H_{2} &=&i\overset{\infty }{\underset{l=0}{\sum }}\left[ \frac{d}{d\phi }\Pi
_{l}^{\dagger }\left( \eta ,\phi ,\lambda \right) \right] \Pi _{l}\left(
\eta ,\phi ,\lambda \right) .  \label{eq23}
\end{eqnarray}

Here, $\Pi _{l}\left( \eta ,\phi ,\lambda \right) =\sqrt{\frac{\left( 1-\eta
\right) ^{l}}{l!}}e^{i\phi \left( n-\lambda l\right) }\eta ^{\frac{n}{2}%
}a^{l}$ is the phase-dependent Krause operator, satisfying $\sum \Pi
_{l}^{\dagger }\left( \eta ,\phi ,\lambda \right) \Pi _{l}\left( \eta ,\phi
,\lambda \right) =1$, with $\lambda =0$ and $\lambda =-1$ representing the
photon losses before the phase shifter and after the phase shifter,
respectively. $n=a^{\dag }a$ is the number operator, and $\eta $ is related
to the dissipation factor with $\eta =1$ and $\eta =0$ being the cases of
complete lossless and absorption, respectively. Following the spirit of Ref.
\cite{e4}, we can further obtain the minimum value of $C_{Q}$ by optimizing
over $\lambda $, corresponding to $F_{L}$, i.e., $F_{L}=\min C_{Q}\leq C_{Q}$%
. Simplifying the calculation process allows us to derive the QFI under
photon losses:
\begin{equation}
F_{L}=\frac{4F\eta \left \langle n_{a}\right \rangle }{\left( 1-\eta \right)
F+4\eta \left \langle n_{a}\right \rangle },  \label{eq24}
\end{equation}%
where $F$ is the QFI in the ideal case \cite{a9}.

Next, we further analyze the effects of each parameter on the QFI of the
single-path LSO scheme under photon losses by numerical calculation. Fig. 10
plots the QFI and QCRB as functions of transmittance $\eta ,$ from which it
is observed that the QFI and the QCRB improve with the rising transmittance $%
\eta $, and the single-path LSO can enhance the QFI and the QCRB. The
improved QFI increases with the transmittance $\eta $, while the improved $%
\Delta \phi _{QCRB_{L}}$ decreases with the transmittance $\eta $.

\begin{figure}[tph]
\label{Figure10} \centering%
\subfigure{
\begin{minipage}[b]{0.5\textwidth}
\includegraphics[width=0.85\textwidth]{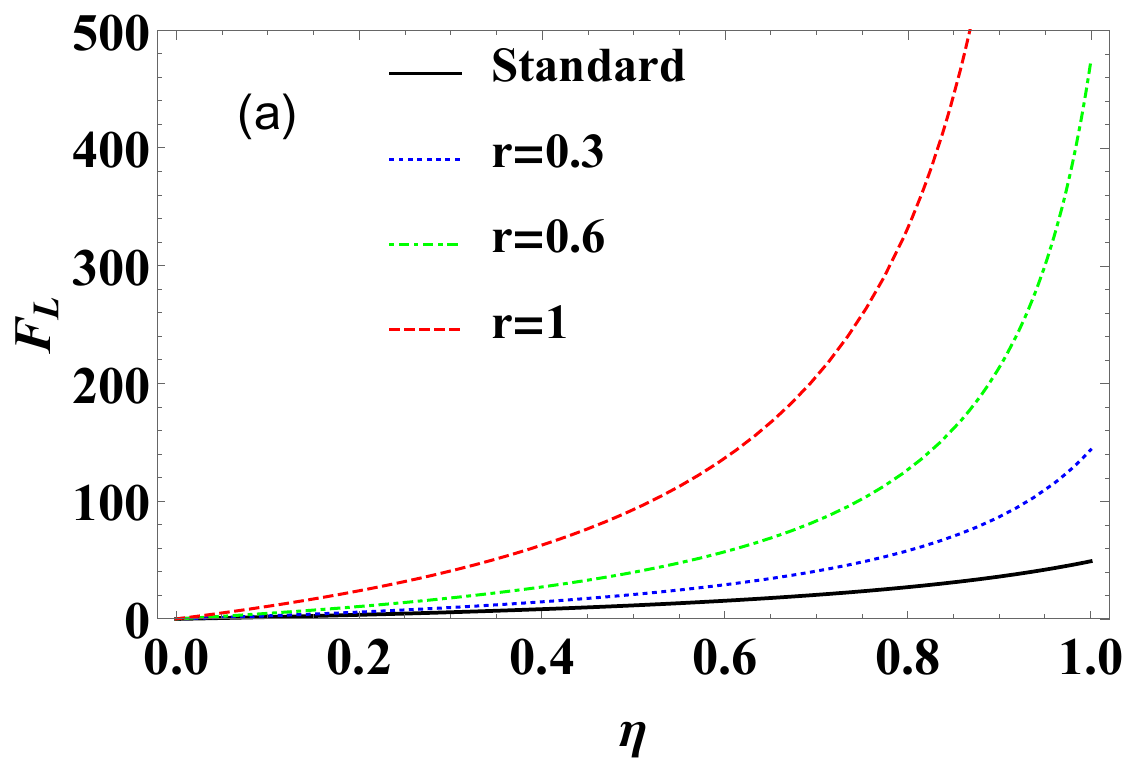}\\
\includegraphics[width=0.85\textwidth]{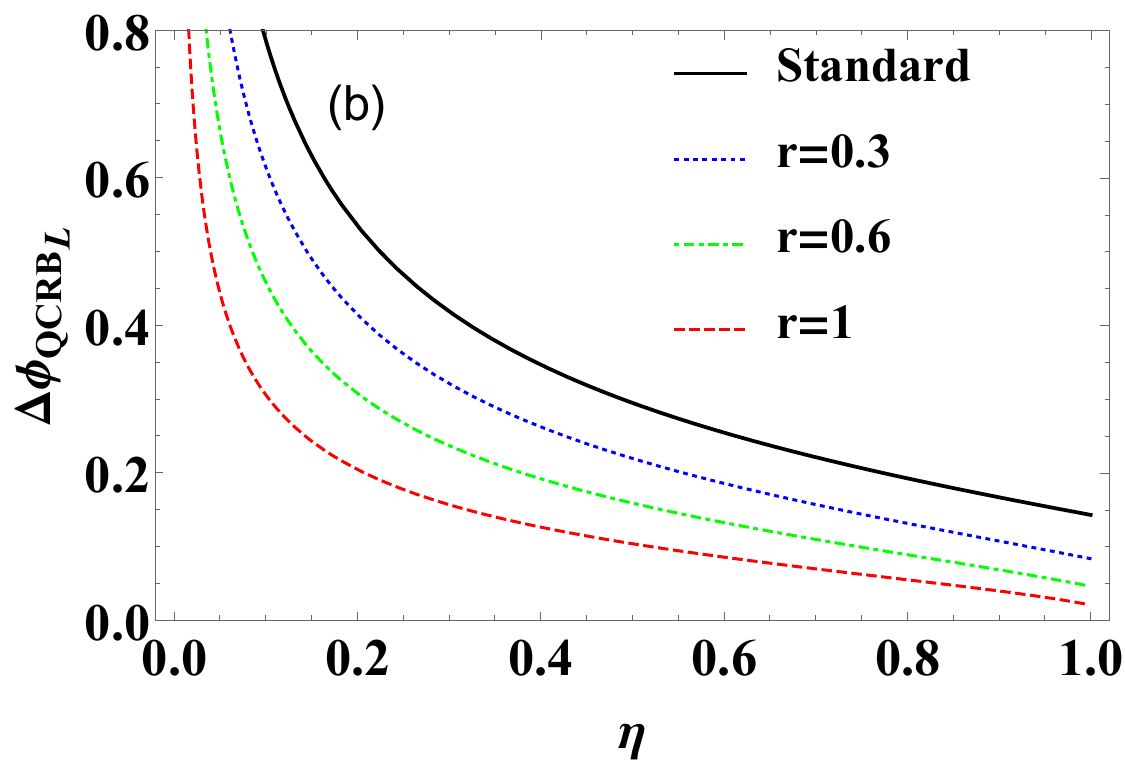}
\end{minipage}}
\caption{The $F_{L}$ and $\Delta \protect \phi _{QCRB_{L}}$ as functions of
transmittance $\protect \eta $, with $g=1$ and $\protect \alpha =1$.}
\end{figure}

Similar to the ideal case, Fig. 11 illustrates the QFI as a function of $g$ (%
$\alpha $) for a given $\alpha $ ($g$), under photon-loss case with $\eta
=0.5$. Similar to Fig. 7, The improvement in QFI increases with increasing
values of the squeezing parameter $r$ and $g$ ($\alpha $).

\begin{figure}[tph]
\label{Figure11} \centering%
\subfigure{
\begin{minipage}[b]{0.5\textwidth}
\includegraphics[width=0.85\textwidth]{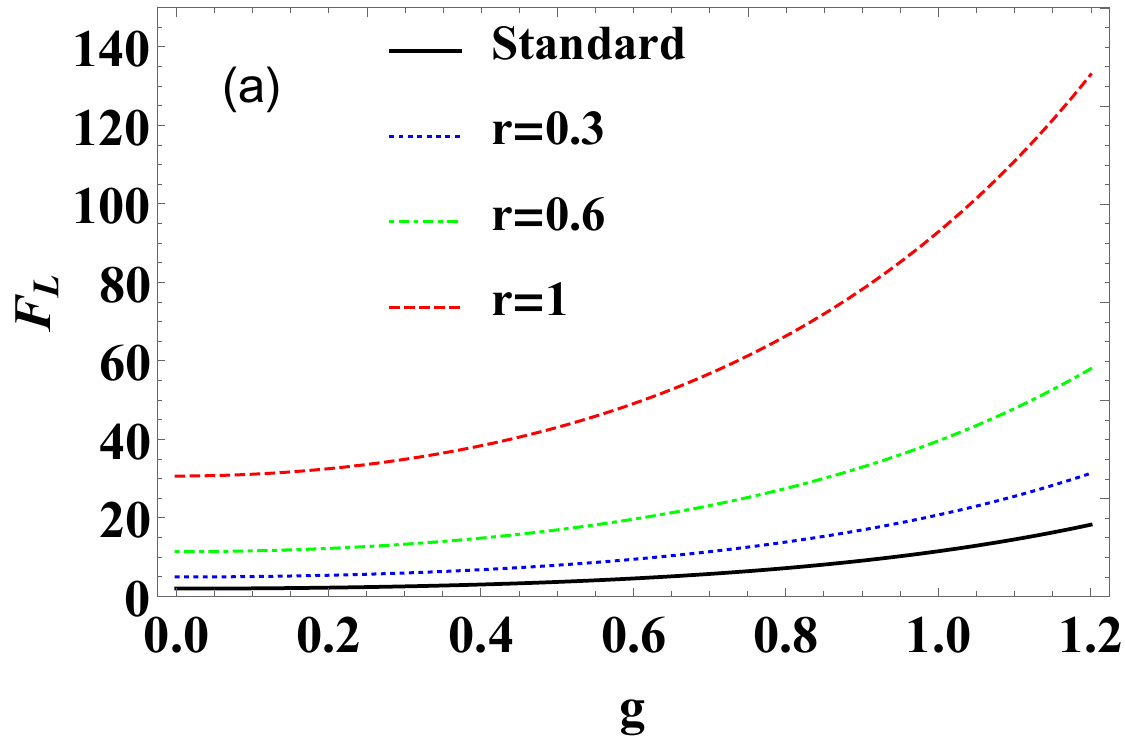}\\
\includegraphics[width=0.85\textwidth]{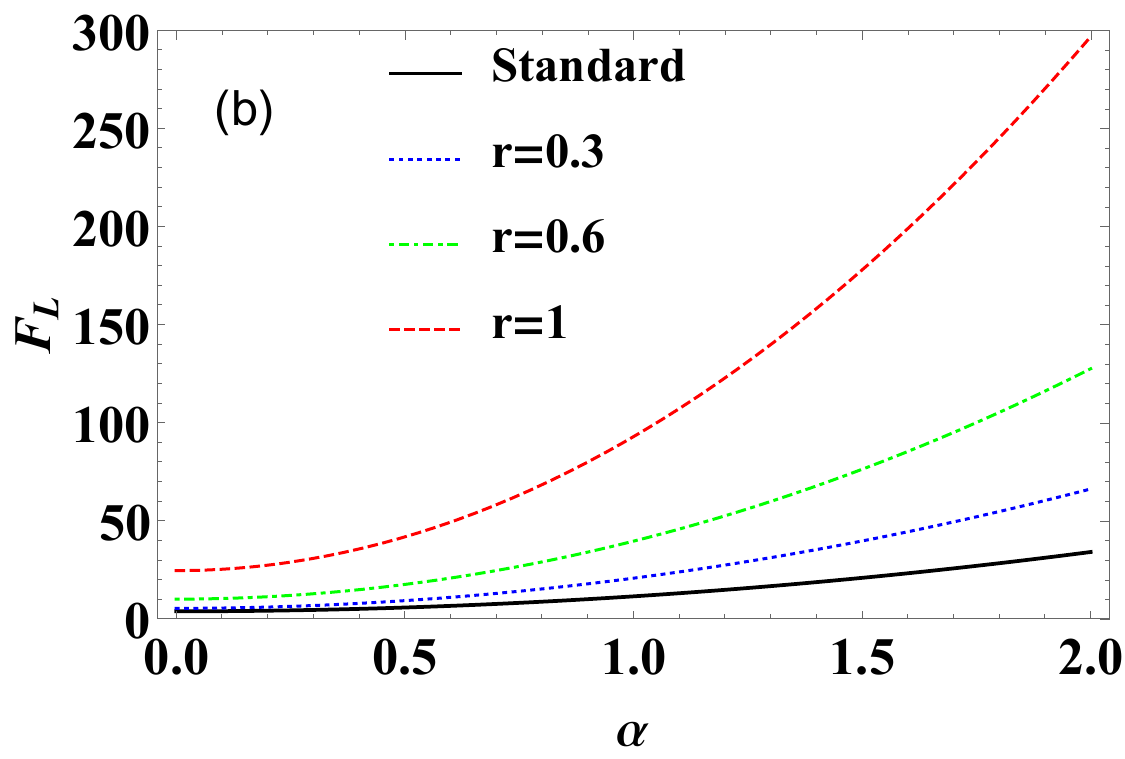}
\end{minipage}}
\caption{(a) The $F_{L}$ as a function of $g$, with $\protect \alpha =1$ \
and $\protect \eta =0.5$. (b) The $F_{L}$ as a function of $\protect \alpha $,
with $g=1$ \ and $\protect \eta =0.5$.}
\end{figure}

\section{Conclusion}

In this paper, we have analyzed the effects of the single-path LSO scheme on
the phase sensitivity and the QFI in both ideal and photon-loss cases.
Additionally, we have investigated the effects of the squeezing parameter $r$
of the single-path LSO scheme, the gain coefficient $g$ of the OPAs, the
amplitude $\alpha $ of the coherent state\ and the transmittance $T_{k}$ of
the BSs on the performance of the system. Through analytical comparison, we
have verified that the single-path LSO scheme can improve the measurement
accuracy of the SU(1,1) interferometer and enhance the robustness against
internal and external photon losses. Throughout the entire range of $T_{k}$,
internal losses have a more significant impact on the system's phase
sensitivity than external losses. This finding indicates that reducing
internal photon losses can be crucial for improving overall performance when
optimizing the design of a quantum interferometer.

Gaussian operations are extensively utilized in quantum optics and
information processing due to their ease of implementation and high energy
efficiency. They enable effective manipulation of quantum states with
relatively straightforward experimental setups, making them suitable for a
wide range of applications. In contrast, non-Gaussian operations, while
capable of generating a rich variety of quantum resources, present
significant challenges. They are typically more complex to implement, often
requiring sophisticated experimental techniques and precise control over
system parameters. Additionally, these operations tend to be more
energy-intensive, which can limit their practicality in certain scenarios.
Our study highlights the potential of Gaussian operations to enhance the
performance of quantum measurement and information processing systems.

In summary, the single-path LSO scheme is crucial in mitigating both
internal and external photon losses in the SU(1,1) interferometer, thereby
effectively improving the accuracy of quantum measurements. The Gaussian
operation not only improves measurement reliability but also offers new
insights and methods for future quantum technology applications.

\begin{acknowledgments}
This work is supported by the National Natural Science Foundation of China (Grants No. 11964013 and No. 12104195) and the Jiangxi Provincial Natural Science Foundation (Grants No. 20242BAB26009 and 20232BAB211033), Jiangxi Provincial Key Laboratory of Advanced Electronic Materials and Devices (Grant No. 2024SSY03011), as well as Jiangxi Civil-Military Integration Research Institute (Grant No. 2024JXRH0Y07).
\end{acknowledgments}\bigskip

\textbf{APPENDIX A : THE PHASE SENSITIVITY WITH THE SINGLE-PATH LSO SCHEME}%
\bigskip

In this appendix, we give the calculation formulas of the phase sensitivity
with single-path LSO as follows:
\begin{equation}
\Delta \phi =\frac{\sqrt{\left \langle \Psi _{out}^{1}\right \vert \left(
a^{\dagger }+a\right) ^{2}\left \vert \Psi _{out}^{1}\right \rangle -\left
\langle \Psi _{out}^{1}\right \vert \left( a^{\dagger }+a\right) \left \vert
\Psi _{out}^{1}\right \rangle ^{2}}}{|\partial \left \langle \Psi
_{out}^{1}\right \vert \left( a^{\dagger }+a\right) \left \vert \Psi
_{out}^{1}\right \rangle /\partial \phi |}.  \tag{A1}
\end{equation}%
Here, the output state $\left \vert \Psi _{out}^{1}\right \rangle $ is given
by Eq. (\ref{eq4}), so the expectations related to the phase sensitivity in
the single-path LSO scheme are specifically calculated as follows \cite{c6}:
\begin{align}
& \left \langle \Psi _{out}^{1}\right \vert \left( a^{\dagger }+a\right)
\left \vert \Psi _{out}^{1}\right \rangle  \notag \\
& =\left \langle \psi _{in}\right \vert U_{S_{1}}^{\dagger }S_{a}^{\dagger
}[U_{\phi }^{\dagger }U_{B_{1}}^{\dagger }U_{S_{2}}^{\dagger
}U_{B_{2}}^{\dagger }\left( a^{\dagger }+a\right)  \notag \\
& \times U_{B_{2}}U_{S_{2}}U_{B_{1}}U_{\phi }]S_{a}U_{S_{1}}|\psi
_{in}\rangle  \notag \\
& =\sqrt{T_{1}T_{2}}\left( e^{i\phi }Q_{1,0,0,0}+e^{-i\phi
}Q_{0,1,0,0}\right) \cosh g  \notag \\
& +\sqrt{T_{2}}\left( Q_{0,0,0,1}+Q_{0,0,1,0}\right) \sinh g,  \tag{A2}
\end{align}%
and
\begin{align}
& \left \langle \Psi _{out}^{1}\right \vert \left( a^{\dagger }+a\right)
^{2}\left \vert \Psi _{out}^{1}\right \rangle  \notag \\
& =\left \langle \psi _{in}\right \vert U_{S_{1}}^{\dagger }S_{a}^{\dagger
}[U_{\phi }^{\dagger }U_{B_{1}}^{\dagger }U_{S_{2}}^{\dagger
}U_{B_{2}}^{\dagger }\left( a^{\dagger }+a\right) ^{2}  \notag \\
& \times U_{B_{2}}U_{S_{2}}U_{B_{1}}U_{\phi }]S_{a}U_{S_{1}}|\psi
_{in}\rangle  \notag \\
& =T_{1}T_{2}\left( 2Q_{1,1,0,0}+e^{2i\phi }Q_{2,0,0,0}+e^{-2i\phi
}Q_{0,2,0,0}\right) \cosh ^{2}g  \notag \\
& +T_{2}\left( 2Q_{0,0,1,1}+2+Q_{0,0,2,0}+Q_{0,0,0,2}\right) \sinh ^{2}g
\notag \\
& +2T_{2}\sqrt{T_{1}}e^{i\phi }\left( Q_{1,0,0,1}+Q_{1,0,1,0}\right) \sinh
g\cosh g  \notag \\
& +2T_{2}\sqrt{T_{1}}e^{-i\phi }\left( Q_{0,1,1,0}+Q_{0,1,0,1}\right) \sinh
g\cosh g+1,  \tag{A3}
\end{align}%
where
\begin{align}
&Q_{x_{1},y_{1},x_{2},y_{2}}  \notag \\
& =\left \langle \psi _{in}\right \vert U_{S_{1}}^{\dagger }S_{a}^{\dagger
}\left( a^{\dagger x1}a^{y1}b^{\dagger x2}b^{y2}\right) S_{a}U_{S_{1}}|\psi
_{in}\rangle  \notag \\
& =\frac{\partial ^{x_{1}+y_{1}+x_{2}+y_{2}}}{\partial \lambda
_{1}^{x_{1}}\partial \lambda _{2}^{y_{1}}\partial \lambda
_{3}^{x_{2}}\partial \lambda _{4}^{y_{2}}}\{e^{w_{4}}\}|_{\lambda
_{1}=\lambda_{2}=\lambda_{3}=\lambda_{4}=0},  \tag{A4}
\end{align}%
with%
\begin{align}
w_{1}& =\frac{1}{2}\left( \lambda _{1}^{2}+\lambda _{2}^{2}\right) \cosh
r\sinh r+\lambda _{1}\lambda _{2}\sinh ^{2}r  \notag \\
& -\left( \lambda _{1}\cosh r\sinh g+\lambda _{2}\sinh r\sinh g\right)
\notag \\
& \times (\lambda _{3}\cosh g-\lambda _{2}\cosh r\sinh g  \notag \\
& -\lambda _{1}\sinh r\sinh g)  \notag \\
& -\lambda _{4}\sinh g(\lambda _{2}\cosh r\cosh g  \notag \\
& +\lambda _{1}\sinh r\cosh g-\lambda _{3}\sinh g),  \tag{A5} \\
w_{2}& =\lambda _{1}\cosh r\cosh g+\lambda _{2}\sinh r\cosh g  \notag \\
& -\lambda _{4}\sinh g,  \tag{A6} \\
\text{\ }w_{3}& =\lambda _{2}\cosh r\cosh g+\lambda _{1}\sinh r\cosh g
\notag \\
& -\lambda _{3}\sinh g,  \tag{A7} \\
w_{4}& =w_{1}+w_{2}\alpha ^{\ast }+w_{3}\alpha .  \tag{A8}
\end{align}

\end{document}